**Hydrated Cable Bacteria Exhibit Protonic Conductivity Over Long Distances**

*Bradley G. Lusk\*, Sheba Morgan, Shawn P. Mulvaney, Brandon Blue, Sam W. LaGasse, Cory D. Cress, J.T.Bjerg, Woo K. Lee, Brian J. Eddie, Jeremy T. Robinson\**

B.G. Lusk, Dr.\*
Science the Earth, Mesa, AZ 85201, United States of America
NRC Postdoctoral Fellow residing in the Chemistry Division, U.S. Naval Research Laboratory, Washington, DC 20375, United States of America
\*ScienceTheEarth@Gmail.com
Orcid ID: https://orcid.org/0000-0002-3094-805X

S. Morgan
HCBU student residing in the Chemistry Division, U.S. Naval Research Laboratory, Washington, DC 20375, United States of America

W. K. Lee, Dr., S. P. Mulvaney, Dr.
Chemistry Division, U.S. Naval Research Laboratory, Washington, DC 20375, United States of America

J.T. Robinson, Dr.\*, C.D. Cress, Dr.
Electronics Sciences and Technology Division, U.S. Naval Research Laboratory, Washington, DC 20375, United States of America
\*Jeremy.T.Robinson10.civ@us.navy.mil

S.W. LaGasse, Dr.
Laboratory for Physical Sciences, 8050 Greenmead Dr, College Park, MD 20740, United States of America

B. Blue, Dr.
NREIP student residing in the Electronics Sciences and Technology Division, U.S. Naval Research Laboratory, Washington, DC 20375, United States of America

B.J. Eddie, Dr.
Center for Bio/Molecular Science and Engineering, U.S. Naval Research Laboratory, Washington, DC 20375, United States of America

J.T. Bjerg, PhD
Postdoctoral researcher, Center for Electromicrobiology (CEM), Section for Microbiology, Department of Biology, Aarhus University, Aarhus C, Denmark

**Significance Statement**

This manuscript documents protonic conductivity across cable bacteria. The discovery of protonic conductivity in cable bacteria provides a putative scaffold through which protons may be transported on the surface of bacteria through the sediment, or perhaps to other organisms. However, despite these hypotheses, the evolutionary benefit of this phenomenon, its role in environmental settings, and its role in microbial interaction remains unknown. The observation of protonic conductivity in cable bacteria, coupled with the development of a transfer printing technique that enables microorganisms to be stamped with protodes, opens up the possibility to assess the presence of protonic conductivity across the surface of other microorganisms and materials, and potentially build bioprotonic devices.

**Abstract**

This study presents the direct measurement of proton transport along filamentous *Desulfobulbaceae*, or cable bacteria. Cable bacteria are filamentous multicellular microorganisms that have garnered much interest due to their ability to serve as electrical conduits, transferring electrons over several millimeters. Our results indicate that cable bacteria can also function as protonic conduits because they contain proton wires that transport protons at distances > 100 µm. We find that protonic conductivity ($\sigma_P$) along cable bacteria varies between samples and is measured as high as $114 \pm 28$ µS cm$^{-1}$ at 25 °C and 70% relative humidity (RH). For cable bacteria, the protonic conductance ($G_P$) and $\sigma_P$ are dependent upon the RH, increasing by as much as 26-fold between 60% and 80% RH. This observation implies that proton transport occurs via the Grotthuss mechanism along water associated with cable bacteria, forming proton wires. In order to determine $\sigma_P$ and $G_P$ along cable bacteria, we implemented a protocol using a modified transfer-printing technique to deposit either palladium interdigitated protodes (IDP), palladium transfer length method (TLM) protodes, or gold interdigitated electrodes (IDE) on top of cable bacteria. Due to the relatively mild nature of the transfer-printing technique, this method should be applicable to a broad array of biological samples and curved materials. The observation of protonic conductivity in cable bacteria presents possibilities for investigating the importance of long-distance proton transport in microbial ecosystems and to potentially build biotic or biomimetic scaffolds to interface with materials via proton mediated gateways or channels.

**Intro**

Proton conductors are materials or biomaterials that allow positively charged protons ($H^+$) to pass through them as the primary charge carriers. Protonic conductivity ($\sigma_P$) has been observed in a broad spectrum of biotic materials including: ampullae of Lorenzini (AoL) jelly in marine animals[1], chitosan[2,3], collagen[4], keratin sulfate[5,6], reflectin[7,8], melanin[9], and bovine serum albumin[10]. Nevertheless, measuring protonic conductivity across the exterior of bacterial cells remains elusive because bacteria placed over protodes, the proton-carrying equivalent of an electrode, demonstrate poor and inconsistent contact. Furthermore, virtually all biological materials are sensitive to high temperature (> 60 °C), organic solvents, high vacuum (< $10^{-6}$ torr), and UV radiation, thus limiting the utility of conventional semiconductor processing techniques to form metal contacts onto bacterial cells. To address these challenges, we have developed a modified transfer-printing technique to stamp palladium (Pd) protodes and gold (Au) electrodes on top of cable bacteria and Nafion® microwires with minimal perturbation, thereby creating more consistent contact with the cells (**Figure 1a-e**).

Filamentous *Desulfobulbaceae*, or cable bacteria, are multicellular bacteria that form electrical conduits within sediment[11,12]. The metabolism of cable bacteria relies on the oxidation of sulfur in anoxic sediment and the transfer of electrons over several centimeters through the sediment to water containing dissolved oxygen[13,14,15,16,17]. During this process, cable bacteria within the sediment generate electrons and protons via the oxidation of hydrogen sulfide ($H_2S$) to sulfate ($SO_4^{2-}$) coupled with the reduction of $O_2$ to $H_2O$, as shown in **Equation 1** and **Equation 2**.

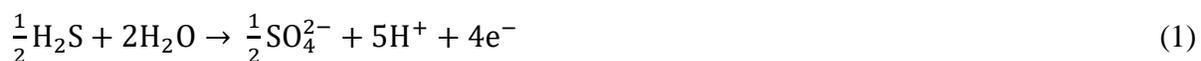

$$\tfrac{1}{2}H_2S + 2H_2O \rightarrow \tfrac{1}{2}SO_4^{2-} + 5H^+ + 4e^- \tag{1}$$

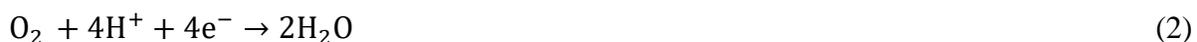

$$O_2 + 4H^+ + 4e^- \rightarrow 2H_2O \tag{2}$$

As shown in Equation 1 and Equation 2, the generation of protons via the oxidation of $H_2S$ is coupled to the consumption of protons during the reduction of $O_2$ for the production of $H_2O$. Previous data shows that, in the presence of cable bacteria, pH within the sediment decreases and the pH near the sediment-water interface becomes slightly basic, while in the absence of cable bacteria, there is no change in pH in the sediment or the sediment-water interface[12,16]. These data suggest that cable bacteria in the sediment may transport protons along their surface to the oxic zone (**Figure 1f**).

Protonic conductivity is measured using palladium protodes- the proton-carrying equivalent of an electrode- consisting of a protonic source and drain made of palladium hydride ($PdH_x$). As shown in **Equation 3**, Pd is an ideal material to use for protodes since it readily adsorbs $H_2$ gas ($PdH_{ads}$) to form a reservoir of $H^+$ ions, resulting in $PdH_{ads}$ splitting into palladium hydride ($PdH_x$), protons, and electrons[1,6,8,18].

$$H_2 + 2Pd \rightarrow 2PdH_{ads} \leftrightarrow 2Pd + 2H^+ + 2e^- \qquad (3)$$

Once a voltage is applied across the protode, protons are injected onto the protonically conductive material, in this case cable bacteria, while electrons corresponding to the number of injected protons are measured as they travel through an external circuit[19,1] (Figure 1e).

Protonic conductivity can result from protons traveling along chains of hydrogen bonds between water and hydrophilic residues via the Grotthuss mechanism[19,20,21], forming what are commonly referred to as proton wires[5,6]. The Grotthuss mechanism is a model for the process of protons diffusing through the network of hydrogen bonds associated with water molecules or other materials via the formation and cleavage of covalent bolds associated with adjacent molecules. Proton wires are formed when water molecules associate with hydrophilic surfaces, typically charged residues in biotic materials including hydroxyl anions ($OH^-$), hydronium cations ($H_3O^+$), or carboxylates that then function as pathways for protonic movement[22,23,24,25,26]. Along proton wires, hydrogen bonds exchange with covalent bonds, resulting in the transfer of protons between adjacent molecules[21,27] (Figure 1f). Given the necessity of water to form these proton wires, protonic conductivity is proportional to the hydration state of the material, which is indirectly measured using relative humidity[28,29].

The hypothesis that cable bacteria may play a role in proton transport led the authors to perform direct measurements of protonic conductivity on the surface of non-living cable bacteria to determine their ability to transport protons over distances up to > 100 μm. The finding of protonic conductivity in cable bacteria is significant since it provides a putative scaffold through which protons may be transported on the surface of bacteria through the sediment, or perhaps to other organisms as is the case with direct interspecies electron transfer (DIET)[30]. However, despite these hypotheses, the evolutionary benefit of this phenomenon, its role in environmental settings, and its role in microbial interaction remains unknown. The observation of protonic conductivity in cable bacteria, coupled with the development of a transfer printing technique that enables microorganisms to be stamped with protodes, opens up the possibility to assess the presence of protonic conductivity across the surface of other microorganisms and materials, and potentially build bioprotonic devices.

**Results**

**Modified transfer-printing technique to adhere protodes directly to the exterior surface of filamentous cable bacteria**.

Prior to applying palladium (Pd) protodes or gold (Au) electrodes to cable bacteria using the modified transfer printing technique, cable bacteria were identified by observing the characteristic ridge structures on their surface using atomic force microscopy (AFM) (Figure 1b). Cable bacteria were determined to be non-viable using a LIVE/DEAD BacLight Bacterial Viability Kit (Invitrogen, USA) and imaged using Confocal Laser Scanning Microscopy (CLSM) (**Figure S1**). As shown in Figure 1c, in order to adhere Pd interdigitated protodes (IDP) – protodes containing 6 gaps of equal length (10 µm), Pd transfer length method (TLM) protodes – protodes containing gaps of various lengths, and Au interdigitated electrodes (IDE) to cable bacteria, a modified transfer printing technique was developed that occurs in four steps: (i) the cable bacteria were dispersed onto a PDMS film on a $SiO_2$/Si substrate; (ii) the protodes or electrodes were adhered to a second PDMS film; (iii) the cable bacteria and protodes were spatially aligned in an optical microscope and contacted such that the cable bacteria span the fingers of the protode; and (iv) the protodes or electrodes were released and the second PDMS film was removed, leaving behind the metallized cable bacteria devices that can be tested at different relative humidities (RH) in an environmental chamber. Figure 1a shows a schematic of cable bacteria stamped with a protode and Figure 1d shows cable bacteria stamped with a TLM protode.

Once samples were introduced to the chamber, a mix of 10% hydrogen gas ($H_2$) and 90% nitrogen gas ($N_2$) was bubbled through DI $H_2O$ at varying rates to adjust the RH of the chamber. As shown in Figure 1e, in the chamber, the Pd absorbs the $H_2$, forming $PdH_x$. As a voltage ($V_{SD}$) is applied across the source (S) and drain (D), the protons ($H^+$) flow through the cable bacteria from the S to the D and electrons (e-) flow through a separate circuit to maintain electroneutrality. Since gold (Au) does not absorb $H_2$ and is not protonically conductive, cable bacteria stamped with Au interdigitated electrodes (IDE) will not transfer $H^+$. Furthermore, since the electrical conductivity of cable bacteria is inhibited by the presence of $O_2$[13], protonic conductivity could be measured after exposing cable bacteria to atmospheric conditions, which is required for the modified transfer printing technique.

**Direct measurement of protonic conductance ($G_P$) along exterior surface of filamentous cable bacteria**.

To determine whether cable bacteria are protonically conductive, this study measured the transport of protons along cable bacteria using Pd interdigitated protodes (IDP) and Pd transfer length method (TLM) protodes. Proton conducting biotic materials should have measurably different current responses in the presence of a humidified environment, a proton source (e.g., $H_2$), and either proton-conductive or electron-conductive contacts. **Figure 2a** shows the current-voltage (*I-V*) response from linear sweep voltammetry (LSV) for cable bacteria stamped with an IDP (Figure 1d) at 80% RH with either 0% or 10% $H_2$. Figure 2a shows a 3.8-5.3 ×** (for all error values reported in the manuscript: ** indicates $p < 0.01$ and * indicates $p < 0.05$.) increase in total conductance ($G_T$) of cable bacteria at 80% RH in the presence of 10% $H_2$ (56 ± 9 pS at 80% RH: 10% $H_2$) compared to 0% $H_2$ (12 ± 0.06 pS), indicating that cable bacteria are protonically conductive. Replicates from four cable bacteria indicate that protonic conductance ($G_P$) varies by cable bacteria and is 44 ± 6 pS, 229 ± 31 pS, 2.6 ± 0.17 nS, and 179 ± 19 pS respectively at 80% RH: 10% $H_2$. Repeat studies with other cable bacteria indicate a 3.2 ± 1.2 × increase** in $G_T$ across samples in the presence of 10% $H_2$ compared to 0% $H_2$ at 80% RH. (See **Table S1** for results from four cable bacteria stamped with IDP at 80% RH.)

Results from IDPs without cable bacteria are included as a control for conductance that may result from the underlying PDMS layer, ions left over after the rinsing process, or adsorbed water layers on the PDMS surface (Figure 2a). For comparison, IDPs in the presence of 10% $H_2$ and lacking cable bacteria are two orders of magnitude (~100 ×)** less conductive than IDPs with cable bacteria, indicating minimal conductance resulting from the underlying PDMS layer or ions left over after the rinsing process.

In order to control for the possibility that enzymes in the dead cable bacteria may be able to catalyze the oxidation of $H_2$ to protons and electrons, resulting in an electrical current, we employed a proton-blocking Au IDE to verify the extent of electronic conductivity. **Figure 2b** shows the *I-V* response from LSV for cable bacteria stamped with Au IDE (Figure 2c) at 80% RH with either 0% or 10% $H_2$. Since Au is a proton-blocker[1,8], cable bacteria stamped with an IDE will only show increased conductance and current in the presence of $H_2$ if it is a source of electrons. Figure 2b shows that cable bacteria are minimally electronically conductive under the conditions measured in this experiment, as indicated in previous experiments[13,12,14]. However, on IDEs, no significant difference in $G_T$ of cable bacteria is observed in the presence of 10% $H_2$ (5.3 ± 0.6 pS) compared to 0% $H_2$ (5.2 ± 0.5 pS). This observation indicates that $H_2$ is not a source of electrons and that the increase in $G_T$ observed for cable bacteria in the presence of 10% $H_2$ on IDP is the result of $H_2$ providing a source of protons. This observation was repeated using multiple cable bacteria (see Table S1 for IDE results from three cable bacteria stamped with Au IDE).

Results from an IDE without cable bacteria are included as a control for conductance that may result from the underlying PDMS layer or ions left over after the rinsing process (Figure 2b). In addition, IDEs in the presence of 10% $H_2$ and lacking cable bacteria are two orders of magnitude (~100 ×)[**] less conductive than IDE with cable bacteria indicating minimal conductivity resulting from the underlying PDMS layer or ions left over after the rinsing process.

In order to confirm that the increase in $G_T$ in the presence of 10% $H_2$ on IDP results from proton transport, cable bacteria were exposed to 10% deuterium gas ($D_2$) instead of 10% $H_2$, thus substituting protons ($H^+$) with deuterium ions ($D^+$)[31]. As shown in **Figure 2d**, since $D^+$ are transported with lower mobility than $H^+$, the kinetic isotope effect decreases conductance compared to $H^+$ loading on the same IDP, causing a decreased response in the *I-V* curve[32,33]. Based on the slope of the *I-V* curve in Figure 2d, the $G_T$ is 20 ± 4 pS at 75% RH: 10% $H_2$, 5.3 ± 0.3 pS at 75% relative humidity using deuterium dioxide (DH): 10% $D_2$, and 0.5 ± 0.3 pS at 75% DH: 0% $H_2$. The $G_T$ for cable bacteria in the presence of 10% $D_2$ is 27-37%[**] the $G_T$ for cable bacteria in the presence of 10% $H_2$ at the same RH, providing strong evidence for proton conduction in cable bacteria[1,32]. (See **Table S2** for results from two cable bacteria stamped with IDP and exposed to $D_2$).

Results from the IDP without cable bacteria are included as a control for conductance that may result from the underlying PDMS layer or ions left over after the rinsing process. In addition, IDPs in the presence of 10% $H_2$ and lacking cable bacteria are close to two orders of magnitude (~80 ×)[**] less conductive than IDE with cable bacteria indicating minimal conductivity resulting from the underlying PDMS layer or ions left over after the rinsing process.

To determine whether protonic conductivity observed across cable bacteria was a consequence of properties inherent in the cable bacteria and not merely water associated with a continuous surface, a control was conducted by applying an IDP to *Microcoleus* – a filamentous bacterium that is known to be non-electronically conductive (**Figure 2f**)[13]. As shown in **Figure 2e**, *Microcoleus* does not exhibit conductivity in the presence or absence of $H_2$. These results indicate that the $G_P$ observed along cable bacteria is likely mediated by charged components of the cable bacteria and not merely the result of cable bacteria serving as a continuous scaffold for water.

**Direct measurements of dependence of relative humidity (RH) on protonic conductivity ($\sigma_P$) along exterior surface of filamentous cable bacteria**.

As shown in **Figure 3a**, due to the variation of protonic conductivity ($\sigma_P$) between cable bacteria, it is difficult to discern a trend related to relative humidity (RH) by analyzing $\sigma_P$ between samples. **Figure 3b** shows that, similar to previous observations of electronic conductivity ($\sigma_E$)[13], the $\sigma_P$ varies between cable bacteria, with a general trend between 324 ± 4.7 nS cm$^{-1**}$ at 60% RH to 6.5 ± 0.4 µS cm$^{-1**}$ at 80% RH, and the highest observed $\sigma_P$ was 114 ± 28 µS cm$^{-1*}$ at 70% RH. (For a list of $\sigma_P$ for 12 cable bacteria, see **Table S1**.)

To control for the variation in $\sigma_P$ between cable bacteria on different IDPs, the RH was adjusted within the same cable bacteria on the same IDP. **Figure 3c** shows that, within the same cable bacteria, there is a positive relationship between $\sigma_P$ and increasing RH. $\sigma_P$ values were acquired by monitoring the *I-V* response of different RH in the presence of 10% H$_2$ (shown in **Figure 3d**), including: 60% (3.5 ± 0.35 nS cm$^{-1}$)$^{**}$, 70% (13.5 ± 1.9 nS cm$^{-1}$)$^{**}$, 75% (59.9 ± 3.5 nS cm$^{-1}$)$^{**}$, and 80% RH: 10% H$_2$ (148 ± 20.7 nS cm$^{-1}$)$^{**}$. Results in Figure 3c are reported as a percentage of the highest observed $\sigma_P$ (3.5 ± 0.5 µσ cm$^{-1}$ at 80% RH: 10% H$_2$). Results from three cable bacteria indicate that $\sigma_P$ is highly dependent upon the RH of cable bacteria, with an increase in $\sigma_P$ 3.4 ± 1.8 ×$^{*}$ between 60% and 70% RH, 2.6 ± 1.7 ×$^{**}$ between 70% and 75% RH, and 2.2 ± 0.9 ×$^{**}$ between 75% and 80% RH; an overall increase in $\sigma_P$ of 15.8 ± 9 ×$^{**}$ between 60% and 80% RH. (**Figure S2** shows influence of RH on the $\sigma_P$ of three cable bacteria.)

**Contact resistance (R$_c$) and specific contact resistivity ($\rho_c$) between cable bacteria, Nafion® microwires, and protodes applied using the modified transfer printing technique**.

Next, the transfer length method (TLM) was employed to determine the contact resistance (R$_c$) and specific contact resistivity ($\rho_c$) of cable bacteria at 75% relative humidity (RH) with either 10% or 0% H$_2$ using the transfer length (L$_T$) and mean diameter of cable bacteria (L$_c$) by stamping them with a protode containing specific gap lengths between the source and drain (L$_g$)[34]. The inverse of the slope obtained from the relationship between L$_g$ and total resistance (R$_T$), determined by linear regression[35] and shown in **Figure 4a**, was used to calculate the total conductivity ($\sigma_T$) and protonic conductivity ($\sigma_P$) of cable bacteria using TLM devices[1,13].

**Figure 4b** shows the gap lengths at which the linear sweep voltammetry (LSV) was performed for each condition.

Results shown in Figure 4a indicate that, in the presence of 10% $H_2$, the $\sigma_T$, $R_c$, and $\rho_c$ of cable bacteria were $0.25 \pm 0.05$ µS cm$^{-1}$, $0.38 \pm 0.06$ TΩ, and $0.71 \pm 0.11$ MΩ-cm$^2$, while at 0% $H_2$, the $\sigma_T$, $R_c$, and $\rho_c$ of cable bacteria were $0.04 \pm 0.004$ µS cm$^{-1}$, $2.5 \pm 0.16$ TΩ, and $4.9 \pm 0.31$ MΩ-cm$^2$ respectively. These measurements reveal that the $\sigma_P$ for cable bacteria under these conditions is $0.22 \pm 0.04$ µS cm$^{-1**}$ and that the $R_c$ accounts for ~25-50% (~$34 \pm 10$%) of $R_T$, indicating that the $\sigma_P$ reported for cable bacteria are conservative estimates.

In order to provide additional confidence in our device fabrication process and our measured $\sigma_P$ of cable bacteria, we employed the transfer length method using Nafion® microwires at 75% RH as a reference material due to its high protonic conductance[19,28]. In the presence of 10% $H_2$, the $\sigma_T$, $R_c$, and $\rho_c$ of Nafion® were $2.6 \pm 1.1$ µS cm$^{-1}$, $1.5 \pm 0.89$ GΩ, and $1.7 \pm 1.0$ KΩ-cm$^2$, while at 0% $H_2$, the $\sigma_T$, $R_c$, and $\rho_c$ of Nafion® were $0.49 \pm 0.19$ µS cm$^{-1}$, $0.16 \pm 0.01$ TΩ, and $3.9 \pm 0.09$ MΩ-cm$^2$. These measurements reveal that the $\sigma_P$ for Nafion® is $2.1 \pm 0.8$ µS cm$^{-1*}$ and that the $R_c$ accounts for approximately 24-51% (~$43 \pm 10$%) of $R_T$, indicating that the $\sigma_P$ reported for Nafion® are conservative estimates.

These results show that cable bacteria have approximately 11%$^{**}$ of the $\sigma_P$ of Nafion® at 75% RH: 10% $H_2$, indicating that Nafion® is about 9.5 × more protonically conductive than cable bacteria under these conditions (Figure 4a). Furthermore, the observation of current production in the absence of 10% $H_2$ for Nafion® ($0.49 \pm 0.19$ µS cm$^{-1}$) likely results from the transfer of protons present in the water[28]. (This hypothesis was verified by stamping Nafion® with a Au TLM; results are shown in **Figure S3**.) Nevertheless, due to the electronic conductivity ($\sigma_E$) of cable bacteria, it is difficult to differentiate the electronic current from the protonic current at 0% $H_2$. For this reason, the reported $\sigma_P$ exclude background current from the 0% $H_2$ condition ($\sigma_{0\%H2}$) for cable bacteria and Nafion®, and thus the $\sigma_P$ reported are conservative estimates.

Since the Grotthuss mechanism plays a significant role in the protonic conductivity of Nafion®[36], the results for a Nafion® microwire should resemble the results of cable bacteria if the proton transport mechanism is the same. **Figure 4d** shows the dependence of RH on the $\sigma_P$ of a Nafion® wire, determined by stamping it with an interdigitated protode (IDP) (**Figure 4c**). Results are reported as a percentage of the highest observed $\sigma_P$ (131.4 ± 7.3 nS cm$^{-1}$ at 80% RH: 10% $H_2$). For Nafion®, $\sigma_P$ values were acquired by monitoring the *I-V* response of different RH in the presence of 10% $H_2$ (**Figure 4e**), including: 60% (5.5 ± 0.16 nS cm$^{-1}$)[**], 70% (11.8 ± 0.5 nS cm$^{-1}$)[**], 75% (58.4 ± 2.4 nS cm$^{-1}$)[**], and 80% RH: 10% $H_2$ (130.4 ± 6.0 nS cm$^{-1}$)[**]. As shown in Figure 4d, the $\sigma_P$ of Nafion® increased by ~2.2 ± 0.6 ×[**] between 60% and 70% RH, ~5.1 ± 1.4 ×[**] between 70% and 75% RH, and ~2.3 ± 0.2 ×[**] between 75% and 80% RH; an overall increase in $\sigma_P$ of ~23.7 ± 3.0 ×[**] between 60% and 80% RH. The results in Figure 3c to Figure 4d show that the $\sigma_P$ in cable bacteria and Nafion® wires share a similar dependence on RH, increasing 15.8 ± 9.0 × and 23.7 ± 3.0 × between 60% and 80% RH respectively, and thus the Grotthuss mechanism is likely a major contributor to $\sigma_P$ in cable bacteria. (Table S3 contains $\sigma_P$ data for two Nafion® wires.)

**Discussion**

Here, we report the first observation of protonic conductivity ($\sigma_P$) along cable bacteria and the protonic contact resistance ($R_c$) between Pd and cable bacteria, supporting the hypothesis that cable bacteria are capable of transporting protons over distances > 100 µm. The highest observed $\sigma_P$ of cable bacteria was $0.11 \pm 0.03$ mS cm$^{-1}$ at 70% RH, putting the $\sigma_P$ of cable bacteria on the same order of magnitude as other protonic materials, including reflectin (0.1 mS at 90% RH)[8], maleic chitosan (0.7 mS at 75% RH)[37], and keratan sulfate ($0.5 \pm 0.11$ mS at 95% RH)[6], and within one order of magnitude of AoL jelly ($2 \pm 1$ mS at 90% RH)[1].

The biological significance of protonic conductivity in cable bacteria is not fully understood since this capability has not yet been confirmed with living cable bacteria and the role active cable bacteria play in the transport of protons in active sediment was not assessed in this manuscript. Furthermore, assuming the $\sigma_P$ observed in dead cable bacteria in this study is similar in live cable bacteria, the $\sigma_P$ offered by cable bacteria residing in sediment is predicted to be ~6 orders of magnitude smaller than the ionic drift current through the pore water[38], indicating that protonic conduction along the surface of cable bacteria may play a minimal role in transporting protons from the anoxic sediment to the oxic zone in the water, thus the putative role of protonic conductivity across the surface of cable bacteria in sediment remains elusive. Nevertheless, the protonic conductivity of cable bacteria is comparable to other protonically conductive biotic materials, indicating that there may be a selective pressure for protonic conductivity in cable bacteria.

Similar to the finding of electrical conductivity in *Geobacter*, the finding of protonic conductivity in cable bacteria opens up a field of research for investigating microbial interactions and communities. For example, when considering the discovery of extracellular electron transfer (EET) in microbial ecosystems, the observation of electric conductivity in *Geobacter*[39] preceded understanding the role of a conductive extracellular matrix[40,41], the mechanisms for EET[42,43,44,45], the composition of the nanowires that compose the extracellular matrix[46], and the discovery that EET plays a pivotal role in direct interspecies electron transfer (DIET)[30] by several decades.

From this initial observation of protonic conductivity in cable bacteria using the modified transfer printing technique described herein, the field can expand to investigate the underlying physiological mechanism(s), determine potential interspecies interactions, and eventually construct accurate predictive models for proton transport in cable bacteria. To better understand the role of proton transport on the surface of cable bacteria in sediment, future studies should focus on investigating protonic conductivity in live isolated cable bacteria and cable bacteria associated with microbial communities. Furthermore, the protonic conductivity of the outer sheath of cable bacteria should be investigated to determine if the sheath plays a role in inferring protonic conductivity as it does with electrical conductivity[13]. Finally, the temperature dependence of protonic conductivity should be investigated to determine the activation energy of proton transport.

Biotic materials that operate as proton conductors via 'proton wires' often increase in $\sigma_P$ as RH is increased since water absorption precedes the formation of hydrogen bonds along which protons are transported[8,37,47]. Furthermore, residues that are aromatic[48], contain carboxyl groups[49], are charged, and are hydrophilic[8], are associated with elevated levels of $\sigma_P$. Since our measurements indicate that cable bacteria are protonically conductive while other filamentous bacteria (ie: *Microcoleus*) are not, we hypothesize that, while in a hydrated state, surface residues contained by cable bacteria may infer protonic conductivity along the length of the cable bacteria via the Grotthuss mechanism by assisting with the formation of proton wires. Future investigations should elucidate the underlying physiology of the scaffold for building the proton wires on the surface of cable bacteria by examining the presence of peptides, proteins, and/or amino acid residues that are aromatic[48], contain carboxyl groups[49], are charged, and/or are hydrophilic[8]. The methods for measuring protonic conductivity in biological samples described herein provide a framework to investigate and characterize protonic conductivity in other microorganisms, filaments, and curved materials. Finally, given that several biological processes are responsive to protons[22], the observation of protonic conductivity in cable bacteria presents possibilities to use these bacteria as scaffolds to build biotic or biomimetic interfaces with materials via proton mediated gateways or channels and, potentially, to investigate the role of microbially mediated proton transport in microbial communities.

**Methods**

**Materials**. $SiO_2$ (100nm)/Si wafers (500-550 µm thick) were purchased from Pure Wafer. Sylgard 184 (Dow, USA) was used for preparing polydimethylsiloxane (PDMS). The spin coater was a WS-400-6NPP-LITE (Laurell, USA). The 3D printer used to construct the transfer arms was a FormLabs 2 equipped with High Temp Resin. The optical microscope was a Nikon equipped with a 50 × objective lens. Linear sweep voltammetry (LSV) was performed using a Keithley 4200 semiconductor analyzer attached to DCM 210 series precision positioners (Cascade Microtech, USA) that were equipped with 0.6 µm PTT-06/4-25 Tungsten needles (Cascade Microtech, USA) and mounted to a grounded M150 stand (Cascade Microtech, USA). All conductivity measurements were acquired with the samples placed inside of a custom-built environmental chamber. Relative humidity (RH) was measured at the source of the gas feed and verified in the environmental chamber using two separate humidity meters (Fisher Scientific, USA). LIVE/DEAD BacLight Bacterial Viability Kit was purchased from Invitrogen (USA) and contained the fluorescent dyes SYTO 9 and propidium iodide. LIVE/DEAD results were obtained using an AxioObserver.Z1/7 LSM 800 (Zeiss, Germany) with a CFI Plan Apochromat Lambda 20X objective (Nikon, Japan). Dynamic atomic force microscopy (AFM) was performed using an Asylum Cypher S (Oxford Instruments, UK).

**Preparing PDMS coated SiO$_2$/Si substrates**. SiO$_2$/Si wafers were cut into ~5 cm x 5 cm squares. PDMS was prepared by mixing a 5:1 ratio of Sylgard 184 silicone elastomer base curing agent and removing air bubbles under vacuum for 30 min. Then a 600 nm thick PDMS layer was deposited onto SiO$_2$/Si wafers by spin coating at room temperature for 10 seconds at 3000 RPM. After spin coating, SiO$_2$/Si wafers coated with PDMS were placed at 60 °C overnight to cure.

**Dispersing cable bacteria onto PDMS coated SiO$_2$/Si substrates**. Live cable bacteria were acquired from sulphidic sediment sampled in Aarhus Bay, Denmark[12]. ~20 ml of sediment was transferred to 36 ml glass test tubes and filled with tap water. The test tubes were then placed in a 500 ml beaker filled with tap water and placed at 4 °C. Alternatively, isolated cable bacteria placed on slides were received from the Center for Electromicrobiology in Aarhus, Denmark. When ready for processing, cable bacteria were separated from sediment by transferring 0.5 ml of sediment into a 1 ml centrifuge tube and spinning at 10,000 RPM for four minutes. After centrifugation, the supernatant was removed, 200 µl of 18 MΩ Millipore water was added to the pellet, and the sample was vortexed for 30 seconds. Then, the sample was spun at 10,000 RPM for four minutes. After centrifugation, 1 µl of supernatant was pipetted onto a PDMS coated SiO$_2$/Si substrate and air dried at room temperature for five minutes. After air drying, excess water was removed using a Kim wipe at the edges of the PDMS film. Next, the PDMS coated SiO$_2$/Si substrate was continuously washed with 18 MΩ Millipore water for 30 seconds and then excess water was removed with a Kim wipe. This wash step was repeated three times. The sample was dried and remaining particulates were removed by spraying the PDMS coated SiO$_2$/Si substrate with high pressure N$_2$.

**Applying micro-protodes and micro-electrodes to cable bacteria**. We used a modified transfer-printing technique to deposit Pd interdigitated protodes (IDP), Pd-TLM, Au interdigitated electrodes (IDE), and Au TLM on top of cable bacteria. First, we defined arrays of metal electrodes onto SiO$_2$/Si substrates using convention photolithography with Shipley 1811 resist, e-beam depositing metal thin films (Pd or Au; P ~5-7 torr; thickness: 75-100 nm), and metal lift-off. In order to reduce the adhesion between the Pd metal and the underlying SiO$_2$/Si substrate, we exposed the Pd metal to H$_2$ gas to form PdH$_x$, this process stresses the metal and thereby induces buckling from the SiO$_2$ surface[50]. After H$^+$ diffuses out of the PdH$_x$, the Pd contacts remain partially unadhered on the SiO$_2$ surface. In order to reduce the adhesion between the Au deposited metal and the underlying SiO$_2$/Si substrate, we exposed samples to an oxygen plasma, which generated 'blisters' or 'bubbles' between the Au electrodes and the SiO2 surface.

We used PDMS and polypropylene carbonate (PPC) stamps, which have different temperature-dependent surface adhesion properties, to "peel off" these unadhered metal contacts and redeposit them onto cable bacteria samples. The stamps were mounted on 3D printed holders, aligned under a microscope to specific Pd or Au contacts of interest, and then brought into contact with the metal electrodes to remove them from the SiO$_2$ surface. Afterwards, these stamps (with metal electrodes) were aligned over cable bacteria on a PDMS coated SiO$_2$/Si substrate using micromanipulators and an optical microscope. The stamp/IDPs were then lowered onto the cable bacteria and released from the stamp by heating to 60 °C for 15-20 minutes. The timeline from isolation to stamping cable bacteria with a protode or electrode was ~1 week.

**Linear Sweep Voltammetry (LSV).** Once cable bacteria were stamped with an IDP, IDE, or TLM they were transferred to an environmental chamber and LSV was conducted using a Keithly 4200 semiconductor analyzer equipped with probes (Cascade Microtech, USA). For LSV measurements, voltage (*V*) was adjusted from -1.0 to 1.0 V in 0.1 V steps to avoid electrolysis. All measurements were repeated at least three times. The environmental chamber was used to control the relative humidity (RH), or DH if deuterium dioxide ($D_2O$) was used, and $H_2$ or $D_2$ concentration by bubbling (via aeration stone) predetermined concentrations of purified 100% $N_2$ gas or 10% $H_2$:90% $N_2$ gas through 18 MΩ Millipore $H_2O$, or 10% $D_2$:90% $N_2$ gas through 99.9% $D_2O$ (Sigma, USA).

**Conductivity Calculation.** Total resistance ($R_T$) and total conductance ($G_T$) were determined by taking the average of the slope of the linear fit for the *I-V* curve and the inverse slope of the linear fit for the *V-I* curve obtained from LSV measurements. For TLM protodes and electrodes, total conductivity ($\sigma_T$) was calculated using **Equation 4**[13]:

$$\sigma_T = \frac{4}{m\pi A_C^2} \qquad (4)$$

Where *m* is the slope of TLM line and $A_c$ is the cross-section surface area of the cable bacteria. When using an IDP or IDE, $\sigma_T$ was calculated using **Equation 5**[13]:

$$\sigma_T = \frac{G_T}{L} \times \frac{L_g}{A_c} \qquad (5)$$

Where *L* is the number of gaps between the electrode/ protode teeth, $L_g$ is the gap length, and $A_c$ is the cross-section surface area of the cable bacteria or Nafion® wire.

The cross-section surface area of the cable bacteria was determined using **Equation 6**[13]:

$$A_c = \frac{\pi D_c^2}{4} \qquad (6)$$

Where $D_c$ is the diameter of the cable bacteria (2.5 µm) or Nafion® wire (6.3 µm). (See supplemental information for discussion on using the cross-sectional area based on the diameter of the cable bacteria rather than using the cross-sectional area based on the diameter of cable bacteria filaments (50 nm)[13].)

In order to directly evaluate the protonic conductivity ($\sigma_P$), the average conductivity of cable bacteria at each RH with 0% $H_2$ ($\sigma_{0\% H2}$) was subtracted from the average conductivity of cable bacteria at each RH with 10% $H_2$ ($\sigma_{10\% H2}$). This subtraction minimizes background current resulting from electronic conductivity ($\sigma_E$) since cable bacteria are electronically conductive[13]. In order to determine statistical significance for the conductance or conductivity of cable bacteria in the presence of 10% $H_2$ vs 0% $H_2$, a paired two tailed student's t-test was conducted with an $\alpha = 0.05$ (indicating 95% certainty) or $\alpha = 0.01$ (indicating 99% certainty). Protonic conductivity ($\sigma_P$) is reported as a mean value and was calculated using **Equation 7**:

$$\sigma_P = \mu_{10\%H2} - \mu_{0\%H2} \tag{7}$$

Where $\mu_{10\% H2}$ is the mean $\sigma_T$ measured from an electrode/ protode in the presence of 10% $H_2$ and $\mu_{0\% H2}$ is the mean $\sigma_T$ measured from an electrode/ protode in the presence of 0% $H_2$. The pooled standard deviation was calculated as shown in **Equation 8**:

$$std_{\sigma P} = \sqrt{\frac{std_{10\%H2}^2 + std_{0\%H2}^2}{2}} \tag{8}$$

Where $std_{\sigma P}$ is the standard deviation of $\sigma_P$, $std_{10\% H2}$ is the standard deviation of $\mu_{10\% H2}$, and $std_{0\% H2}$ is that standard deviation of $\mu_{0\% H2}$. Protonic conductance ($G_P$) was calculated by substituting conductance with conductivity measurements in Equations 7 and 8.

**Contact resistance and specific contact resistivity calculations**. The TLM method was used for calculating the contact resistance ($R_c$) and specific contact resistivity ($\rho_c$)[34]. Since $R_c$ is ½ the y-intercept, $R_c$ was determined using the trend lines in Figure 4. $P_c$ was determined using

**Equation 9**:

$$\rho_c = R_{sh} \times L_T^2 \tag{9}$$

Where $R_{sh}$ is the sheet resistance and $L_T^2$ is the current transfer length. Since $L_T$ is ½ the x-intercept, $L_T$ was determined by extrapolating the trend lines in Figure 4 to the x-axis. $R_{sh}$ was determined using **Equation 10**:

$$R_{sh} = m \times D_c \tag{10}$$

**Confocal Laser Scanning Microscopy (CLSM)**. Cable bacteria on PDMS were stained with LIVE/DEAD BacLight stain diluted in 1X phosphate buffered saline (pH 7.4, ThermoFisher, USA) for 10 minutes, after which they were gently washed by five serial exchanges of the medium. The cable bacteria were immediately imaged using the manufacturer's standard protocol. Three-dimensional images were obtained by combining z-stacks of 18 viewfields.

**Atomic Force Microscopy (AFM)**. After depositing cable bacteria on PDMS coated $SiO_2$/Si substrate, samples were imaged using dynamic AFM at room temperature. Stamped and unstamped cable bacteria were observed to confirm conformation of the IDP and IDE to the cable bacteria.

**Supporting Information**
Supporting Information is available online at https://doi.org/10.1073/pnas.2416008122.

**Acknowledgements**
This work was funded by the Office of Naval Research through base programs at NRL. This research was performed while B.G.L. and S.L. held an NRC Research Associateship Post Doctoral Fellowship at NRL. This research was performed while S.M. was an HCBU student residing in the Chemistry Division at the NRL. Finally, this research was performed while S.P.M. was the section head for surface nanoscience & sensor technology at the NRL. J.J.B.'s portion of the research was funded by The Danish National Research Foundation (Center of Excellence DNRF136) and the Carlsberg Foundation (CF19-0666 and CF21-0409 to J.J.B.).

Published January 13, 2025 in the *Proceedings of the National Academy of Sciences of the United States of America*: https://doi.org/10.1073/pnas.2416008122

**Conflict of interest**
The authors declare no conflict of interest.

**Data, Materials, and Software Availability**
The data supporting the main findings of this study are available in the paper and/or SI Appendix.

**References**

1. Josberger, E., et al. Proton conductivity in ampullae of Lorenzini jelly. Science Advances 2(5), e1600112–e1600112 (2016). https://doi.org/10.1126/sciadv.1600112
2. Fischer, S.A., Dunlap, B.I., & Gunlycke, D. Proton transport through hydrated chitosan-based polymer membranes under electric fields. Journal of Polymer Science. Part B, Polymer Physic, 55(14), 1103–1109 (2017). https://doi.org/10.1002/polb.24361
3. Lee, W., et al. Enhanced protonic conductivity and IFET behavior in individual proton-doped electrospun chitosan fibers. Journal of Materials Chemistry C 7(35), 10833–10840 (2019). https://doi.org/10.1039/c9tc02452b
4. Bardelmeyer, G.H. Electrical conduction in hydrated collagen. I. Conductivity mechanisms. Biopolymers 12, 2289-2302 (1973).
5. Morowitz, H.J. Proton semiconductors and energy transduction in biological-systems. American Journal of Physiology 235(3), R99-R114 (1978). https://doi.org/10.1152/ajpregu.1978.235.3.R99
6. Selberg, J., Jia, M., & Marco Rolandi. Proton conductivity of glycosaminoglycans. PloS ONE 14(3), e0202713 (2019). https://doi.org/10.1371/journal.pone.0202713
7. Kautz, R., et al. Review of Cephalopod-Derived Biopolymers for Ionic and Protonic Transistors. Advanced Materials 30(19), e1704917 (2018). https://doi.org/10.1002/adma.201704917
8. Ordinario, D.D., et al. Bulk protonic conductivity in a cephalopod structural protein. Nature Chemistry 6(7), 596–602 (2014). https://doi.org/10.1038/nchem.1960
9. Wünsche, J., et al. Protonic and Electronic Transport in Hydrated Thin Films of the Pigment Eumelanin. Chemistry of Materials 27(2), 436–442 (2015). https://doi.org/10.1021/cm502939r
10. Amdursky, N., et al. Long-Range Proton Conduction across Free-Standing Serum Albumin Mats. Advanced Materials 28(14), 2692–2698 (2016). https://doi.org/10.1002/adma.201505337
11. Bjerg, J., et al. Long-distance electron transport in individual, living cable bacteria. Proceedings of the National Academy of Sciences of the United States of America 115(22), 5786–5791 (2018). https://doi.org/10.1073/pnas.1800367115
12. Pfeffer, C., et al. Filamentous bacteria transport electrons over centimetre distances. Nature 491(7423), 218–221 (2012). https://doi.org/10.1038/nature11586

13. Meysman, F., et al. A highly conductive fibre network enables centimetre-scale electron transport in multicellular cable bacteria. Nature Communications 10(1), 4120–4120 (2019). https://doi.org/10.1038/s41467-019-12115-7
14. Malkin, S.Y., et al. Natural occurrence of microbial sulphur oxidation by long-range electron transport in the seafloor. The ISME Journal 8(9), 1843–18454 (2014). https://doi.org/10.1038/ismej.2014.41
15. Matturro, B., et al. Cable Bacteria and the Bioelectrochemical Snorkel: The Natural and Engineered Facets Playing a Role in Hydrocarbons Degradation in Marine Sediments. Frontiers in Microbiology 8(MAY), 952–952 (2017). https://doi.org/10.3389/fmicb.2017.00952
16. Meysman, F., Risgaard-Petersen, N., Malkin, S., & Nielsen, L. (2015). The geochemical fingerprint of microbial long-distance electron transport in the seafloor. Geochimica et Cosmochimica Acta, 152I, 122–142. https://doi.org/10.1016/j.gca.2014.12.014
17. Bjerg, J. J. Lustermans, J.J.M., Marshall, I.P.G. et al. Cable bacteria with electric connection to oxygen attract flocks of diverse bacteria. Nature Communications 14:1614 1-8 (2023). https://doi.org/10.1038/s41467-023-37272-8
18. Robinson, P., et al. Electrical and electrochemical characterization of proton transfer at the interface between chitosan and PdHx. Journal of Materials Chemistry. C, Materials for Optical and Electronic Devices 5(42), 11083–11091 (2017). https://doi.org/10.1039/C7TC03215C
19. Miyake, R. & Rolandi, M. Grotthuss mechanisms: from proton transport in proton wires to bioprotonic devices. Journal of Physics. Condensed Matter 28(2), 023001–023001 (2016). https://doi.org/10.1088/0953-8984/28/2/023001
20. Grotthuss, C. J. T. Sur la de´composition de l'eau et des corps qu'elle tient en dissolution a` l'aide de l'e´lectricite´ galvanique. Annales de chimie et de physique 58, 54–73 (1806).
21. Cukierman, S. Et tu, Grotthuss! And other unfinished stories. BBA – Bioenergetics 1757(8), 876–885 (2006). https://doi.org/10.1016/j.bbabio.2005.12.001
22. Silverstein, T. P., The Proton in Biochemistry: Impacts on Bioenergetics, Biophysical Chemistry, and Bioorganic Chemistry. Frontiers in Molecular Biophysics 8(764099) 1-18 (2021). https://doi.org/10.3389/fmolb.2021.764099
23. Eigen, M., & De Maeyer, L. Self-Dissociation and Protonic Charge Transport in Water and Ice. Proceedings of the Royal Society of London. Series A, Mathematical and Physical Sciences (1934-1990) 247(1251), 505–533 (1958). https://doi.org/10.1098/rspa.1958.0208
24. Deng, Y., et al. H+-type and OH–type biological protonic semiconductors and complementary devices. Scientific Reports 3(2), 2481 (2013). https://doi.org/10.1038/srep02481
25. Deng, Y., Helms, B., & Rolandi, M. Synthesis of pyridine chitosan and its protonic conductivity. Journal of Polymer Science Part A: Polymer Chemistry 53(2), 211–214 (2015). https://doi.org/10.1002/pola.27430

26. Lee, J., et al. Proton Conduction in a Tyrosine-Rich Peptide/Manganese Oxide Hybrid Nanofilm. Advanced Functional Materials 27(35), 1702185 (2017). https://doi.org/10.1002/adfm.201702185
27. Fischer, S.A., & Gunlycke, D. Analysis of Correlated Dynamics in the Grotthuss Mechanism of Proton Diffusion. The Journal of Physical Chemistry. B 123(26), 5536–5544 (2019). https://doi.org/10.1021/acs.jpcb.9b02610
28. Eknauth, P., & Vona, M.L.E. Hydration and proton conductivity of ionomers: the model case of Sulfonated Aromatic Polymers. Frontiers in Energy Research 2(50), 1-6 (2014). https://doi.org/10.3389/fenrg.2014.00050
29. Sone, Y. & Simonsson, D. Proton Conductivity of Nafion 117 as Measured by a Four-Electrode AC Impedance Method. Journal of the Electrochemical Society 143(4), 1254–1259 (1996). https://doi.org/10.1149/1.1836625
30. Rotaru AE, Woodard TL, Nevin KP, Lovley DR. Link between capacity for current production and syntrophic growth in Geobacter species. Front Microbiol. 2015 Jul 21;6:744. doi: 10.3389/fmicb.2015.00744.
31. Jin, Q. & Kirk, M.F. pH as a Primary Control in Environmental Microbiology: 1. Thermodynamic Perspective. Frontiers in Environmental Science 6, 1-21 (2018). https://doi.org/10.3389/fenvs.2018.00021
32. Rienecker, S. B., Mostert, A. B., Schenk, G., Hanson, G. R., Meredith, P. Heavy water as a probe of the free radical nature and electrical conductivity of melanin. J. Phys. Chem. B 119, 14994–15000 (2015).
33. Roberts, N. & Northey, H. Proton and deuteron mobility in normal and heavy water solutions of electrolytes. Journal of the Chemical Society, Faraday Transactions 1: Physical Chemistry in Condensed Phases 70, 253-262 (1974). https://doi.org/10.1039/f19747000253
34. Abbas, T. & Slewa, L. Transmission line method (TLM) measurement of (metal/ZnS) contact resistance. International Journal of Nanoelectronics & Materials 8, 111-120 (2015).
35. Andrade, J.M. and Estévez-Pérez, M.G. Statistical comparison of the slopes of two regression lines: A tutorial. Analytica Chimica Acta 838, 1–12 (2014). https://doi.org/10.1016/j.aca.2014.04.057
36. Kuwertz, R., Kirstein, C., Turek, T., & Kunz, U. Influence of acid pretreatment on ionic conductivity of Nafion® membranes. Journal of Membrane Science, 500, 225–235. (2016). https://doi.org/10.1016/j.memsci.2015.11.022
37. Zhong, C., Deng, Y., Roudsari, A. F., Kapetanovic, A., Anantram, M. P., & Rolandi, M. A polysaccharide bioprotonic field-effect transistor. Nature Communications, 2(1), 476–476. (2011). https://doi.org/10.1038/ncomms1489
38. Risgaard-Petersen, N., Kristiansen, M., Frederiksen, R. B., Dittmer, A. L., Bjerg, J. T., Trojan, D., Schreiber, L., Damgaard, L. R., Schramm, A., & Nielsen, L. P. Cable Bacteria in Freshwater Sediments. Applied and Environmental Microbiology, 81(17), 6003–6011. (2015). https://doi.org/10.1128/aem.01064-15

39. Lovley DR, Phillips EJ. Novel mode of microbial energy metabolism: organic carbon oxidation coupled to dissimilatory reduction of iron or manganese. Appl Environ Microbiol. 1988 Jun;54(6):1472-80. doi: 10.1128/aem.54.6.1472-1480.1988.
40. Reguera, G., McCarthy, K. D., Mehta, T., Nicoll, J. S., Tuominen, M. T., & Lovley, D. R. (2005). Extracellular electron transfer via microbial nanowires. In Nature (Vol. 435, Issue 7045, pp. 1098–1101). Springer Science and Business Media LLC. https://doi.org/10.1038/nature03661
41. Malvankar, N., Vargas, M., Nevin, K. et al. Tunable metallic-like conductivity in microbial nanowire networks. Nature Nanotech 6, 573–579 (2011). https://doi.org/10.1038/nnano.2011.119
42. Strycharz-Glaven S. M., Snider R. M., Guiseppi-Elie A., Tender L. M. On the electrical conductivity of microbial nanowires and biofilms. Energy & Environmental Science. 2011;4(11):4366–4379. doi: 10.1039/c1ee01753e.
43. Vargas M, Malvankar NS, Tremblay PL, Leang C, Smith JA, Patel P, Snoeyenbos-West O, Nevin KP, Lovley DR. Aromatic amino acids required for pili conductivity and long-range extracellular electron transport in Geobacter sulfurreducens. mBio. 2013 Mar 12;4(2):e00105-13. doi: 10.1128/mBio.00105-13. Erratum in: MBio. 2013;4(2):e00210-13.
44. Xiao K., Malvankar N. S., Shu C., Martz E., Lovley D. R., Sun X. Low energy atomic models suggesting a pilus structure that could account for electrical conductivity of geobacter sulfurreducens pili. Scientific Reports. 2016;6(1):p. 23385. doi: 10.1038/srep23385.
45. Shu C, Zhu Q, Xiao K, Hou Y, Ma H, Ma J, Sun X. Direct Extracellular Electron Transfer of the Geobacter sulfurreducens Pili Relevant to Interaromatic Distances. Biomed Res Int. 2019 Nov 11;2019:6151587. doi: 10.1155/2019/6151587.
46. Wang, F., et al. Structure of microbial nanowires reveals stacked hemes that transport electrons over micrometers. Cell 177, 361–369 (2019). Doi: 10.1016/j.cell.2019.03.029
47. Wraight, C. Chance and design—Proton transfer in water, channels and bioenergetic proteins. B–A - Bioenergetics 1757(8), 886–912 (2006). https://doi.org/10.1016/j.bbabio.2006.06.017
48. Yardeni, J.L., Amit, M., Ashkenasy, G., & Ashkenasy, N. Sequence dependent proton conduction in self-assembled peptide nanostructures. Nanoscale 8(4), 2358–2366 (2016). https://doi.org/10.1039/c5nr06750b
49. Lee, J., et al. Proton Conduction in a Tyrosine-Rich Peptide/Manganese Oxide Hybrid Nanofilm. Advanced Functional Materials 27(35), 1702185 (2017). https://doi.org/10.1002/adfm.201702185
50. Josberger, D.E., Deng, Y., Sun, W., Kautz, R., Rolandi, M. Two-Terminal Protonic Devices with Synaptic-Like Short-Term Depression and Device Memory. Advanced Materials 26(29), 4986–4990 (2014). https://doi.org/10.1002/adma.201400320

**Figure 1.** a) Schematic of cable bacteria stamped with a palladium (Pd) protodes. b) Atomic force microscopy (AFM) image of cable bacteria junction showing characteristic ridge structures on the surface of adjacent cells. c) Overview of how Pd interdigitated protodes (IDP), transfer length method (TLM) protodes, and Au interdigitated electrodes (IDE) were adhered to the cable bacteria using a modified transfer printing technique. d) Image of cable bacteria fragment stamped with a TLM protode – a protode containing gaps of various lengths (scale bar = 50 µm). e) Overview of the experimental design for probing cable bacteria that have been stamped with a Pd protode or a gold (Au) electrode on a PDMS substrate. f) Theoretical framework for how putative proton wires on the surface of cable bacteria mediate the transfer of protons between adjacent water molecules via the Grotthuss mechanism (the presence of carboxyl groups and the ability of cable bacteria to transport protons from the oxic zone to the anoxic zone have not been confirmed).

**Figure 2.** a) *I-V* curve for linear sweep voltammetry (LSV) performed on cable bacteria spanning six ~10 µm gaps on a Pd interdigitated protode (IDP) at 80% relative humidity (RH) and in the presence of either 10% (red) or 0% (black) $H_2$. A control is included showing the *I-V* curve for a Pd IDP without cable bacteria. b) *I-V* curve for cable bacteria spanning six ~10 µm gaps on a Au IDE at 80% RH and in the presence of either 10% (red) or 0% (black) $H_2$. A control is included showing the *I-V* curve for a Au interdigitated electrode (IDE) without cable bacteria. c) A picture of the Au IDE stamped on cable bacteria is shown (scale bar = 50 µm). d) *I-V* curve for cable bacteria spanning six ~10 µm gaps on a Pd IDP at 75% RH and in the presence of either 10% (red) $H_2$ or 0% (black) $H_2$ or 10% (purple) deuterium gas ($D_2$). e) *I-V* curve for *Microcoleus* spanning a ~100 µm gap on a Pd IDP at 80% RH and in the presence of either 10% (red) or 0% (black) $H_2$. f) A picture of the Pd IDP stamped on *Microcoleus* is shown (scale bar = 50 µm).

**Figure 3.** a) Average protonic conductivity ($\sigma_P$) from different cable bacteria samples as a function of relative humidity (RH) in the presence of 10% $H_2$. b) Histogram of $\sigma_P$ recorded from samples in (a). c) Bar graph showing $\sigma_P$ within cable bacteria samples as a function of RH, normalized to percent difference of maximum observed $\sigma_P$ ($3.5 \pm 0.5$ µσ cm$^{-1}$) at 80% RH and in the presence of 10% $H_2$. Results were collected from cable bacteria spanning a Pd interdigitated protode (IDP) with six ~10 µm gaps, respectively. d) *I-V* curve recorded for cable bacteria at various RH levels.

**Figure 4.** a) Total resistance ($R_T$) versus channel width (gap length) for transfer length measurement (TLM) protodes using either cable bacteria or a Nafion® microwire at 75% relative humidity (RH) and in the presence of 10% (red) or 0% (black) $H_2$. Linear regression of the fits (dashed lines) was used to calculate $\sigma_P$ and determine significance. b) Shows gap lengths for each $R_T$ measurement. Squares represent values for cable bacteria and triangles represent values for Nafion® microwires. Measurements were taken in the presence of 10% (red) or 0% (black) $H_2$. c) A Nafion® microwire stamped with a Pd interdigitated protode (IDP) (scale bar = 100 µm). d) $\sigma_P$ of Nafion® microwire as a function of RH, normalized to percent difference of maximum observed $\sigma_P$ (59.4 ± 3.3 µσ cm$^{-1}$) at 80% RH in the presence of 10% H$_2$. e) *I-V* curves recorded for Nafion® microwires at various RH.

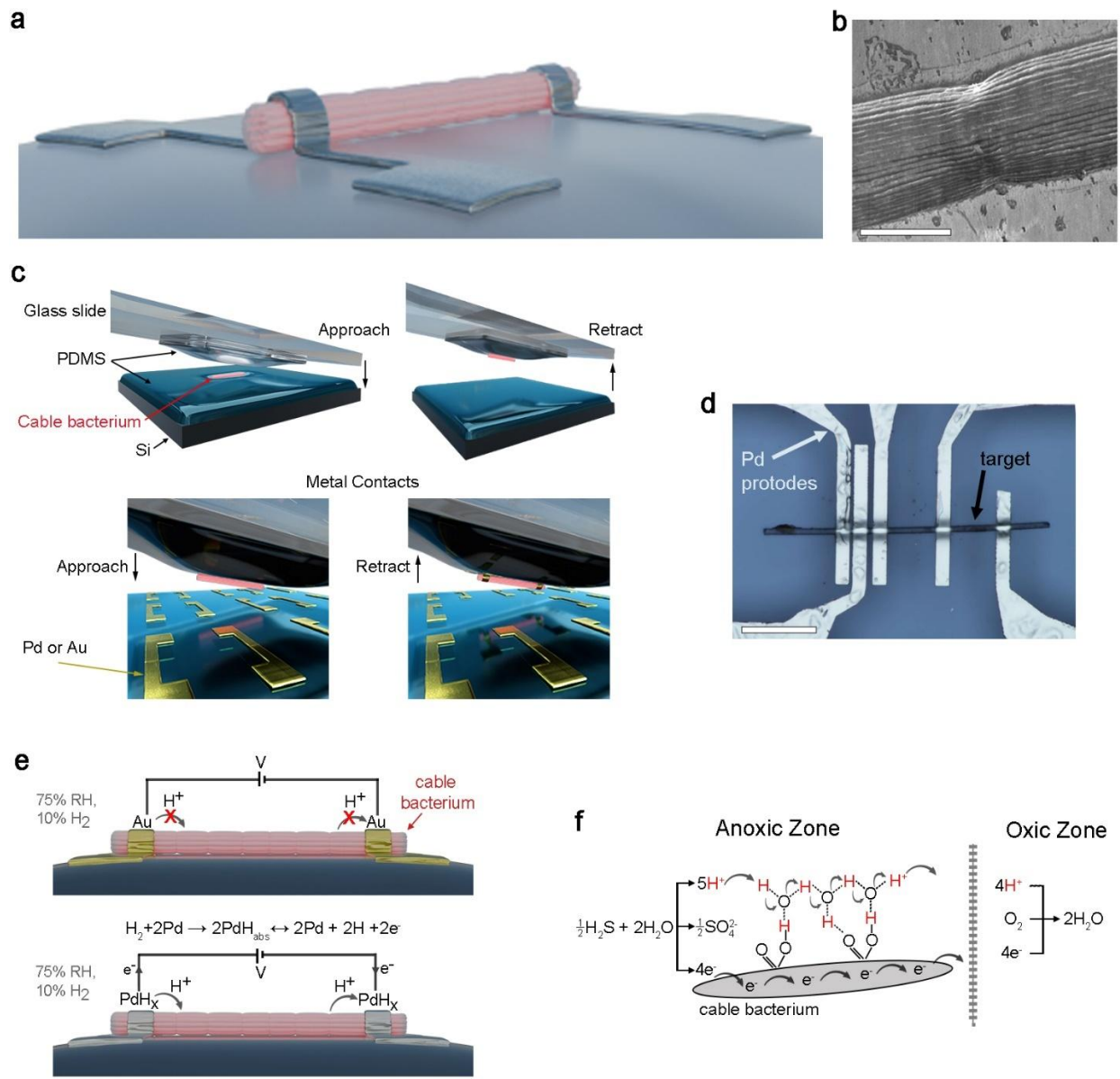

**Figure 1.**

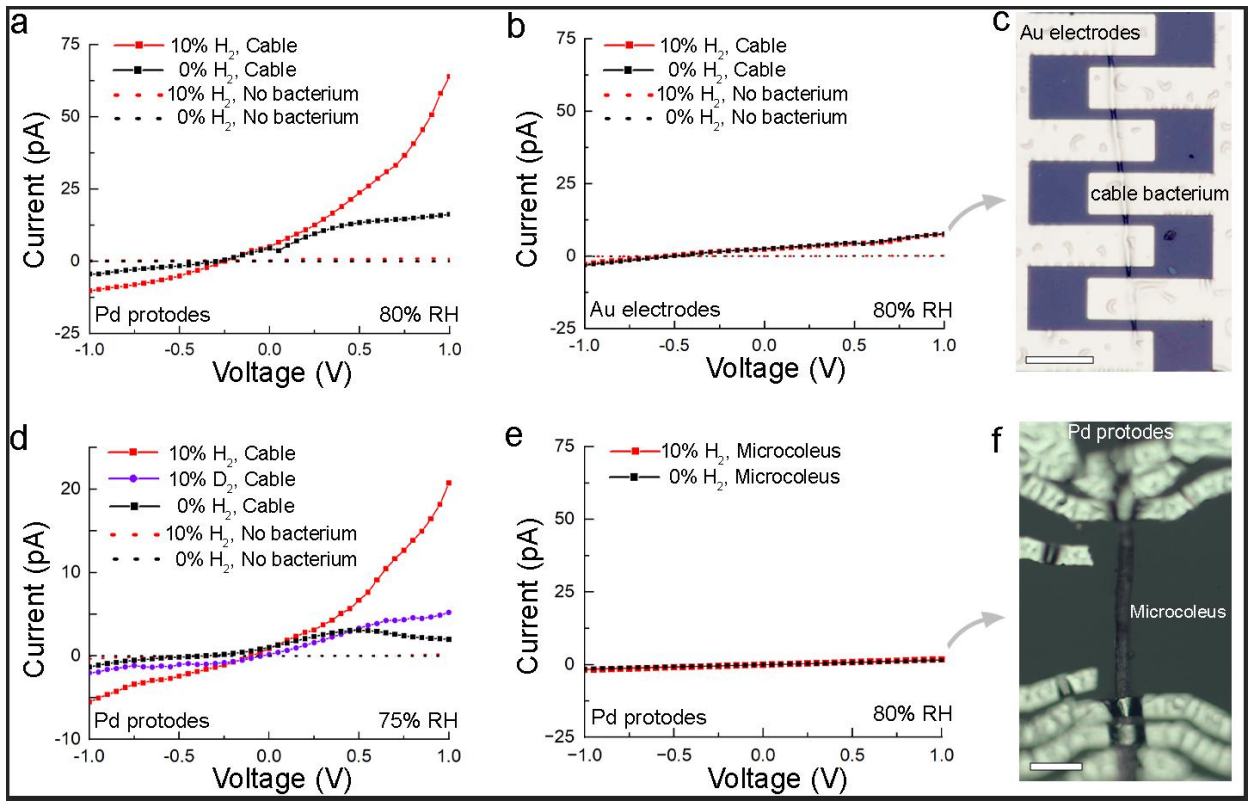

**Figure 2.**

Published January 13, 2025 in the *Proceedings of the National Academy of Sciences of the United States of America*: https://doi.org/10.1073/pnas.2416008122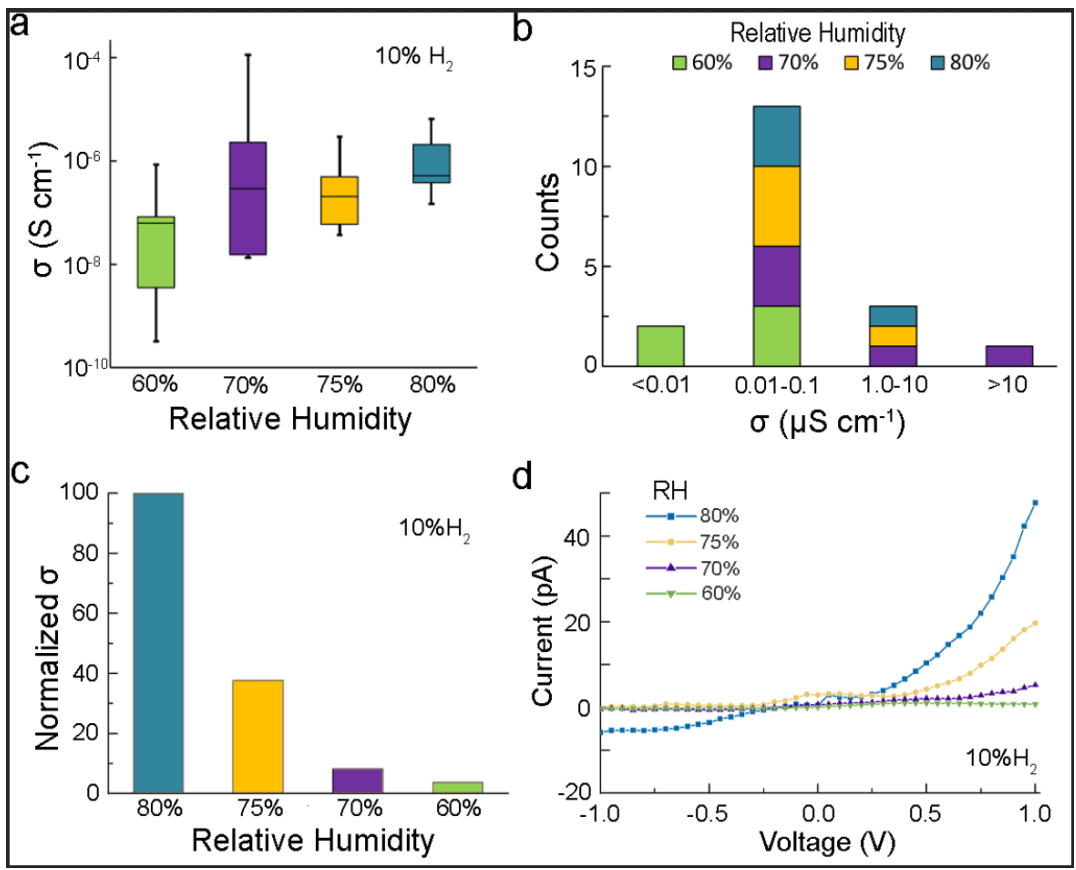

**Figure 3.**

Published January 13, 2025 in the *Proceedings of the National Academy of Sciences of the United States of America*: https://doi.org/10.1073/pnas.2416008122
30

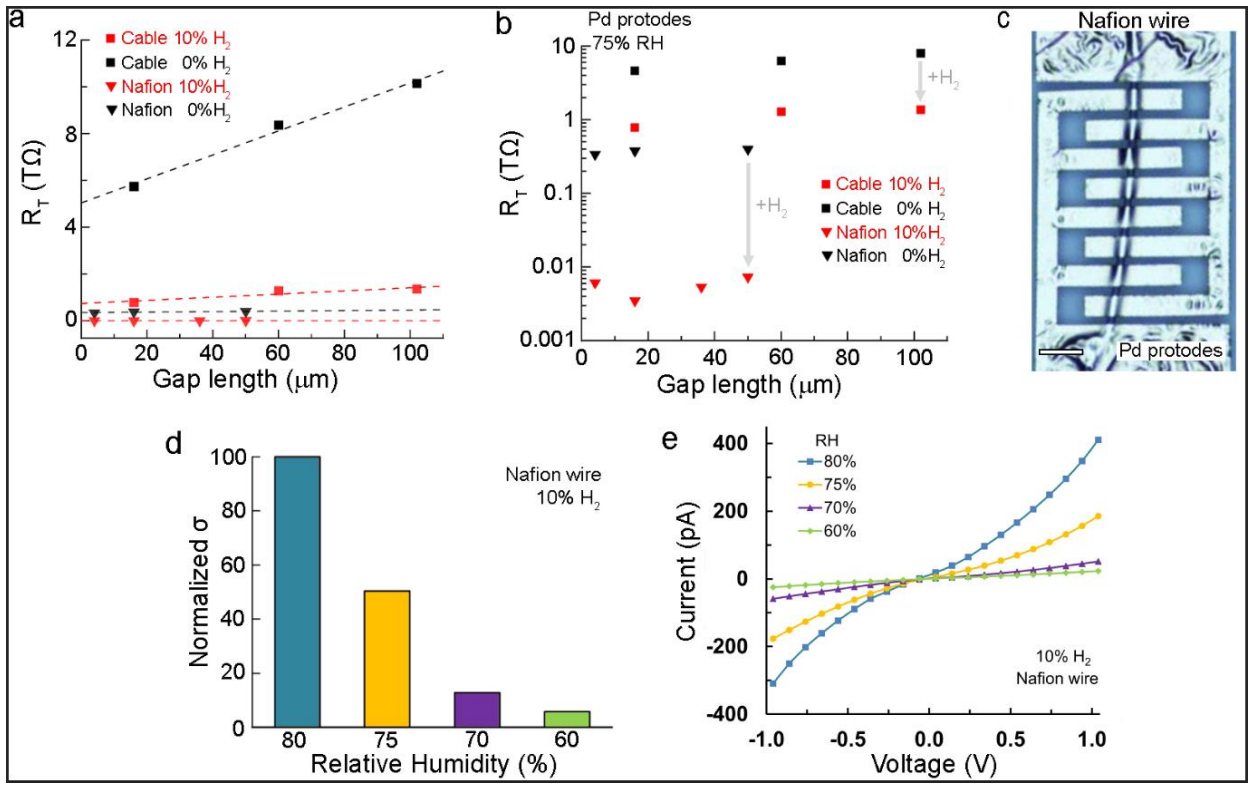

**Figure 4.**